\documentclass[aps,prl,twocolumn,superscriptaddress,amsmath,amssymb]{revtex4}
\usepackage{graphicx}
\usepackage{epstopdf}
\usepackage{amsmath}
\usepackage{color}
\usepackage[normalem]{ulem}
\usepackage{graphicx}
\usepackage{subfigure}
\usepackage{float}
\usepackage{mathrsfs}
\usepackage{bbold}
\usepackage{psfrag}
\usepackage{mathcomp}
\usepackage{verbatim}
\usepackage{multirow}
\usepackage{diagbox}
\usepackage[colorlinks,citecolor=blue]{hyperref}

\usepackage{simplewick}

\def\ba{\begin{eqnarray}}
\def\ea{\end{eqnarray}}

\begin{document}

\title{Mott Insulator-Density Ordered Superfluid Transition and ``Shamrock Transition" in a Frustrated Triangle Lattice}

\author{Ce Wang}
\affiliation{School of Physics Science and Engineering, Tongji University, Shanghai, 200092, China}
\affiliation{Institute for Advanced Study, Tsinghua University, Beijing, 100084, China}

\author{Yu Chen}
\email{ychen@gscaep.ac.cn}
\affiliation{Graduate School of China Academy of Engineering Physics, Beijing, 100193, China}
\date{\today}

\begin{abstract}
Density order is usually a consequence of the competition between long-range and short-range interactions. Here we report a density ordered superfluid emergent from a homogeneous Mott insulator due to the competition between frustrations and local interactions. This transition is found in a Bose-Hubbard model on a frustrated triangle lattice with an extra pairing term. Further, we find a quantum phase transition between two different density ordered superfluids, which is beyond the Landau-Ginzburg paradigm. Across this transition, a U(1) symmetry is emergent, while the symmetry in each density ordered superfluid is $Z_2\times Z_3$. Because there emerges a shamrock-like degenerate ground state in parameter space, we call the transition ``shamrock transition". Effective low energy theories are established for the two transitions mentioned above and we find their resemblance and differences with clock models.
\end{abstract}

\maketitle

{\color{blue}\emph{Introduction}} The coexistence of a superfluid and density order is usually thought to be exotic. Since the density order usually means lack of mobility and superfluid means dissipativeless flow.  Closer scrutiny tells us a density-ordered superfluid is a condensation that happens at a finite momentum, therefore density order and superfluid are not completely contradictory. This kind of phase can be achieved by competition between short-range and long-range interactions\cite{supersoild_review,supersoild_theory1,supersoild_theory2,supersoild_theory3,supersoild_experiment1,supersoild_experiment2,supersoild_dipolar, Blakie}, such as in cavity BEC systems\cite{Esslinger-Dicke,RotonSoftening,Hemmerich,BHMcav,FirstOrder} as well as in dipolar gases systems\cite{dipolar18,dipolar19,Pfau19,dipolar21}, or by dispersion modification such as the stripe phase in Bose-Einstein condensate with spin-orbit-coupling\cite{Hui10,Jason11,Yun12,Yun13,Spielman11,SOC,SOC_stripe}. Among these mechanisms, either we need a finite-range interaction or a dispersion modification to locate kinetic energy minimal to a finite momentum state. Now we try to go beyond these two mechanisms.

In this letter, we give an example of density ordered superfluid phase by the competition between frustration and local interactions. To be more specific, we find a homogenous Mott insulator to density wave ordered superfluid (DWSF) transition in a frustrated triangle lattice system with interacting bosons. The only difference in the new model is an extra on-site pairing term. The Bose-Hubbard model on a triangular lattice with positive hopping strength is well studied previously. A connection is found between the frustrated Bose-Hubbard model and the KT phase. A 120-degree chiral order is developed to reconcile the requirement of homogeneous phase by strong local interactions and the sign structure required by frustration, while no density order is presented\cite{Altman14}. It seems that the presence of frustrations and local interactions is not enough for density order, but the situation changes dramatically when we add an on-sie pairing term\cite{Feng19}. A density ordered superfluid is generated to avoid frustration. These on-site pairing terms are natural in Kerr cavity systems\cite{pairhopping_kc1}, exciton-polariton  systems\cite{pairhopping_ep1,pairhopping_ep2} as well as in NMR systems with anti-punching technique\cite{Jiangfeng}. However, the previous work is only focused on the weak interaction region where the density order is generated in presence of any weak pairing term.  Here in this paper, we are going to study the Mott region, and we find that  the homogenous phase is robust against the perturbation of the on-site pairing term. Therefore the DWSF is generated by spontaneous symmetry breaking.
\begin{figure}[t]
\includegraphics[width=8.8cm]{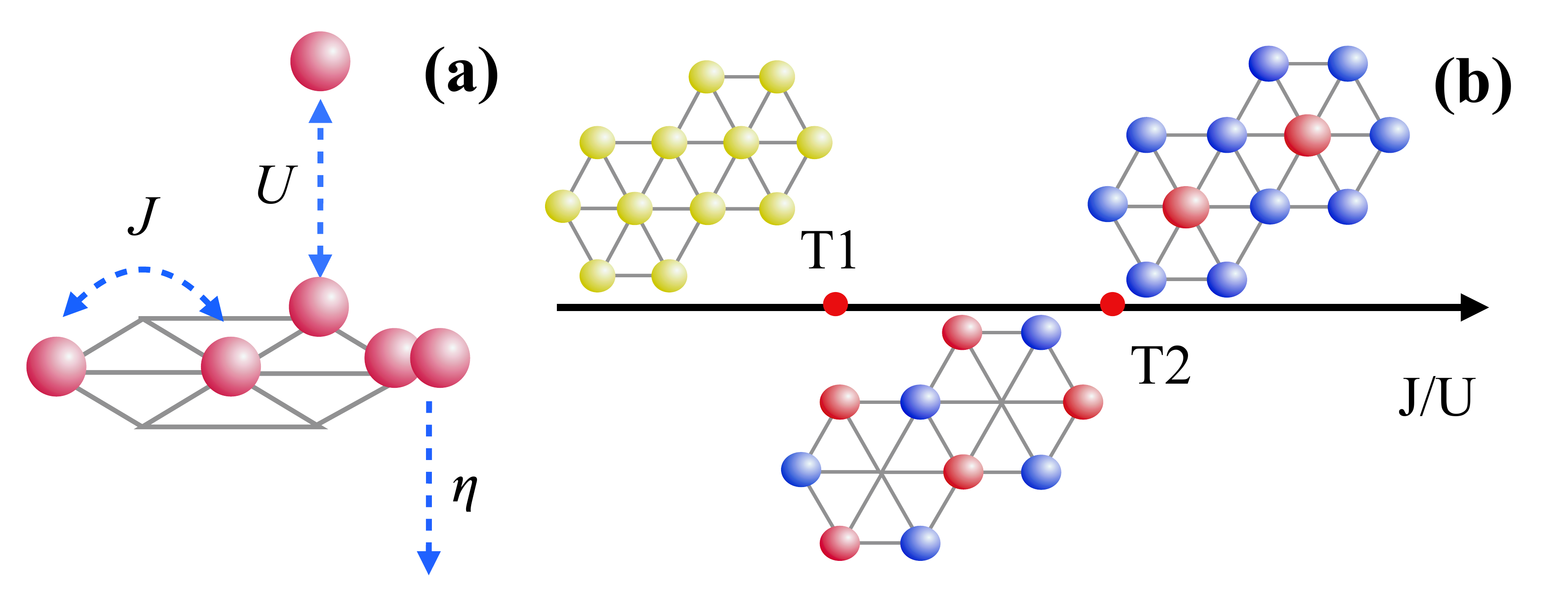}
\caption{(a) Illustration of the model. A Bose-Hubbard model on a triangle lattice with pair generation term. (b) Typical phases for fixed $\eta/U$ with increasing $J/U$. There are Mott insulator, non-chiral supersolid, chiral supersolid phases.}\label{Illu}
\end{figure}

In the following, we will first present our model and then we give the phase diagram by a standard mean-field theory method. We find there are three typical phase transitions as we increase the frustration with fixed pairing energy, the first transition is a Mott insulator-DWSF transition. Distinct with frustrated Bose-Hubbard model without pairing term, this DWSF is non-chiral. The second transition is between two non-chiral DWSFs. In this transition, there is an emergent U(1) symmetry and the low energy minimum shapes as a shamrock in parameter space. For convenience, we call it ``Shamrock transition". The third one is a chiral symmetry broken transition where a chiral ordered DWSF is generated. Here we focus on the first two transitions ( Fig.~\ref{Illu}(b) ) and give the low energy effective theory for them. By these low energy effective theories, we find the mechanism for the generation of DWSF, and we find an emergent U(1) symmetry at the "Shamrock transition". 
This new model supplies us with a new direction to study XY-like models where magnitude mode fluctuations are unlocked. 

{\color{blue}\emph{Model}} Here we consider a Bose-Hubbard model on a triangular lattice with an on-site pairing term,
\ba
\hat{H}&=&\hat{H}_{\rm BH}+\hat{H}_{\rm PR},\label{Eq1}
\ea
where
\ba
\hat{H}_{\rm BH}&=&J\sum_{\langle ij\rangle}(\hat{a}_i^\dag\hat{a}^{}_j+\hat{a}_j^\dag\hat{a}_i)-\sum_i(\mu\hat{n}_i-\frac{U}{2}\hat{n}_i(\hat{n}_i-1)),\nonumber\\
\hat{H}_{\rm PR}&=&-\eta\sum_i(\hat{a}_i^{}\hat{a}_i^{}+\hat{a}_i^\dag\hat{a}_i^\dag).
\ea
Here $\hat{a}_i$ is an annihilation operator of a boson on site $i$ of a triangle lattice, $J>0$ is the positive hopping strength, $\mu$ is the chemical potential, $U>0$ is the on-site interaction strength of bosons and $\eta$ is the local pairing strength. $\hat{n}_i=\hat{a}_i^\dag\hat{a}^{}_i$ is the particle number on site $i$, $\langle ij\rangle$ represents $i$ and $j$ are nearest neighbors.

Before we dive into details of the ground state properties of Eq.~\ref{Eq1}, let us first analyze the symmetry property of this model. One can see there is a translational symmetry on lattice sites. If $\eta=0$, there is a $U(1)$ symmetry, that is the hamiltonian being invariant under transformation $\hat{a}_i\rightarrow \hat{a}_i e^{i\theta}$. It is broken to a $Z_2$ symmetry $\hat{a}_i\rightarrow \pm \hat{a}_i$ when the pairing term is added. In this sense, the superfluid breaks a $Z_2$ symmetry rather than a $U(1)$ symmetry. Here we stress that for the Mott insulator phase, there is still a charge gap even though the particle number is not conserved and the superfluid spontaneously breaks the $Z_2$ symmetry instead of the $U(1)$ symmetry. Although particle number is not conserved in the Mott insulator phase, there is no off-diagonal long-range order between particle annihilation and particle creation. In this sense, the concept of Mott insulator and superfluid is still well-defined. 

\begin{figure}[b]
\includegraphics[width=8.cm]{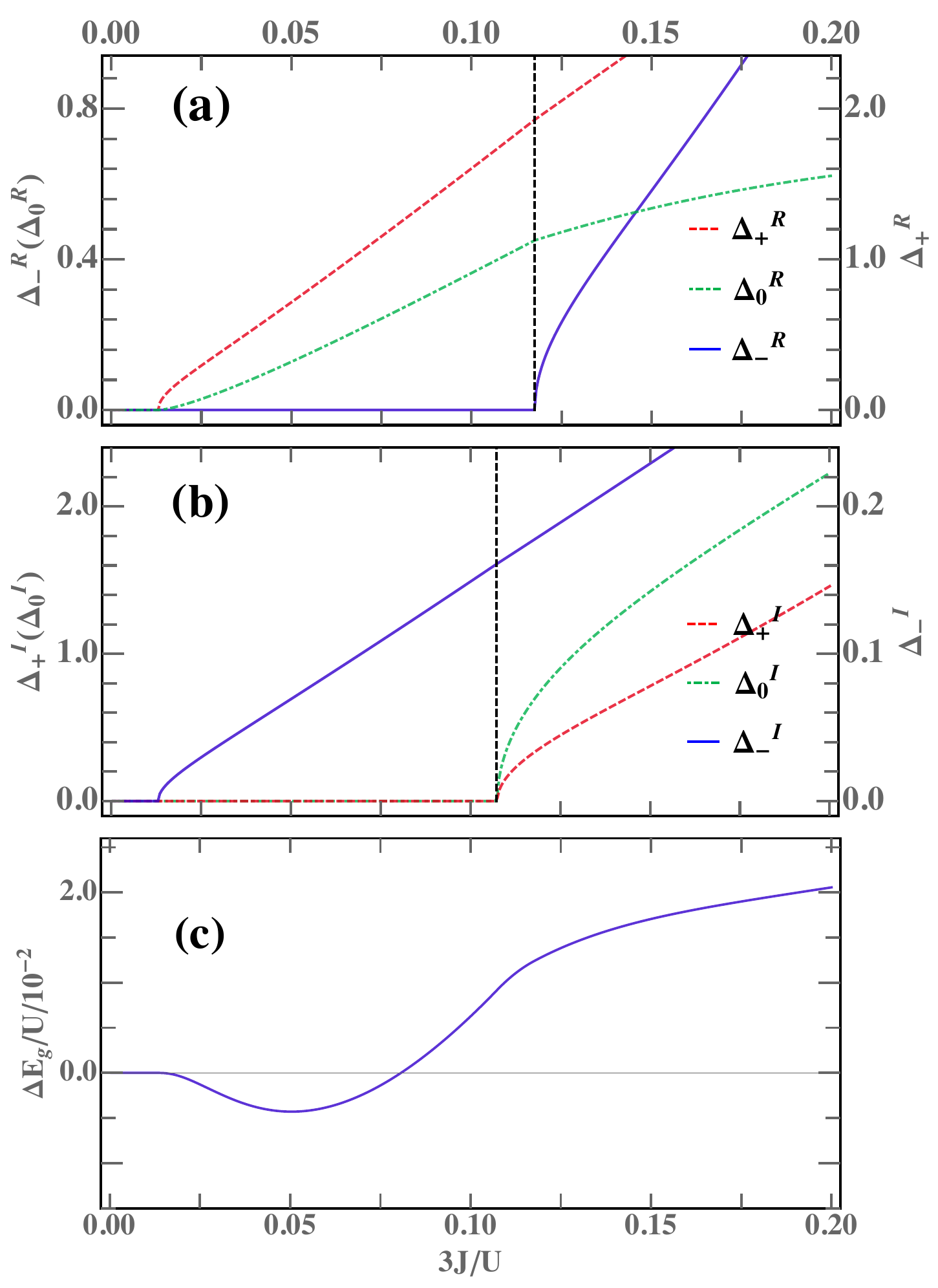}
\caption{ A typical order parameter evolution over $J/U$ when $\eta/U=0.06$ and $\mu/U=2.875$ are fixed. In (a), we show a solution for $\varphi^I= 0$, $\varphi^R\neq0$ (superfluid A). The blue line is for $\Delta_-^R$, the red dashed line is for $\Delta_+^R$ and the green dot-dashed line is for $\Delta_0^R$. In (b), we show another solution for $\varphi^R= 0$, $\varphi^I\neq0$ (superfluid B). The blue line is for $\Delta_-^I$, red dashed line is for $\Delta_+^I$ and green dot-dashed line is for $\Delta_0^R$. Finally, in (c), we show the energy difference $\Delta E_g$ between superfluid A and superfluid B. $\Delta E_g<0$ means that superfluid B phase has a lower energy. }
\label{Order}
\end{figure}
{\color{blue}\emph{Mean Field Theory and Phase Diagram}}
Now we apply a mean field theory to Hamiltonian Eq. ~\ref{Eq1}. First we propose a mean field hamiltonian $\hat{H}_{\rm MF}$ as follows,
\ba
\hat{H}_{\rm MF}&\!=\!&\sum_i \left(\hat{\cal H}_i\!+\!(\varphi_{i-1}^*\!+\!\varphi_{i+1}^*)\hat{a}_i^{}\!+\!(\varphi_{i-1}\!+\!\varphi_{i+1})\hat{a}_i^\dag\right)\\
\hat{\cal H}_i&\!=\!&-\mu\hat{n}_i+\frac{U}{2}\hat{n}_i(\hat{n}_i-1)-\eta(\hat{a}_i\hat{a}_i+\hat{a}_i^\dag\hat{a}_i^\dag),
\ea
where $i\in\{1,2,3\}$ are sites indices in a unit cell with three sites in it, ${\boldsymbol \varphi}=(\varphi_1,\varphi_2,\varphi_3)$ are complex order parameters. The addition and subtraction in $\{1,2,3\}$ are cyclonic, which means $1-1=3$, $3+1=1$.
Now we suppose an eigenstate of $\hat{H}_{\rm MF}$ is $|g_{\boldsymbol\varphi}\rangle$. We take this solution as an ansatz of the ground state of the original hamiltonian. Then we can obtain the ground state energy as
\ba
E_g=\langle g_{\boldsymbol\varphi}|\hat{H}|g_{\boldsymbol\varphi}\rangle.
\ea 

By minimizing ground state energy $E_g$, then we obtain the mean field solution for $\varphi_{i=1,2,3}$ by equations 
\ba
\varphi_i=3J\langle g_{\boldsymbol\varphi}|\hat{a}_i|g_{\boldsymbol\varphi}\rangle.
\ea 
To better describe the order parameter, we find a decomposition as follows,
\ba
{\boldsymbol\varphi}&=&{\boldsymbol\varphi}^R+{\boldsymbol\varphi}^I\nonumber\\
{\boldsymbol\varphi}^R&=&\Delta_0^R\hat{\varphi}_0+\Delta_+^R\hat{\varphi}_++i\Delta_-^R\hat{\varphi}_-\label{Eq:Decomp}\\
{\boldsymbol\varphi}^I&=&i\Delta_0^I\hat{\varphi}_0\ \!+i\Delta_+^I\hat{\varphi}_+-\Delta_-^I\hat{\varphi}_-\nonumber
\ea
where $\hat{\varphi}_0=(1,1,1)/\sqrt{3}$, $\hat{\varphi}_+=(2,-1,-1)/\sqrt{6}$, and $\hat{\varphi}_-=(0,1,-1)/\sqrt{2}$ are three unit vectors. We introduce $\Delta_0=\Delta_0^R+i\Delta_0^I$, $\Delta_\pm=\Delta_\pm^R+i\Delta_\pm^I$, where $\Delta_{\pm}^R$, $\Delta_\pm^I$ and $\Delta_{0}^{R,I}$ are real. A numerical solution can be obtained and we show $\Delta_0$, and $\Delta_\pm$ for different $J/U$ and fixed $\mu/U=2.875$ and $\eta/U=0.06$ in Fig.~\ref{Order}.

\begin{figure}[t]
\includegraphics[width=7.5cm]{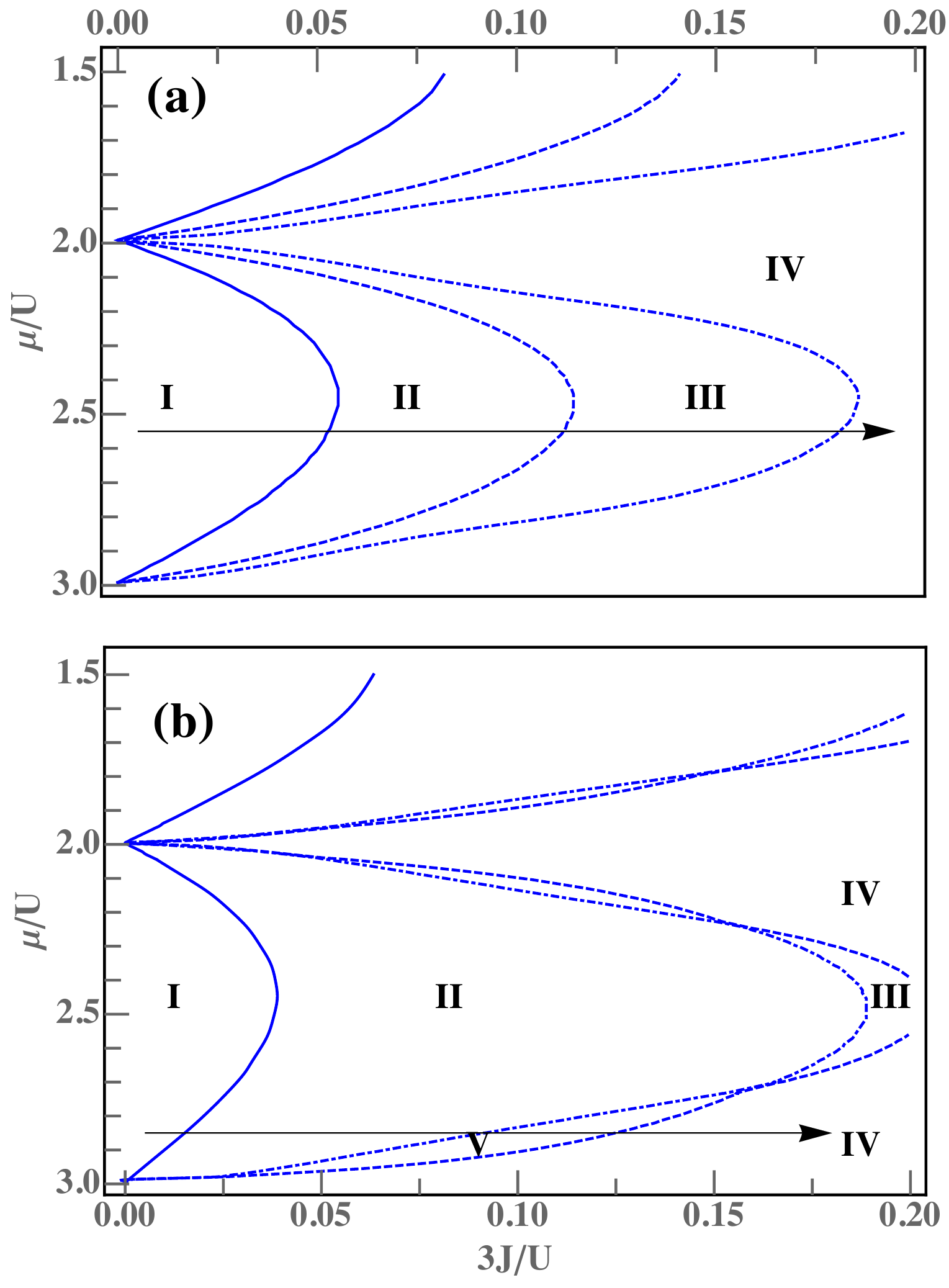}

\caption{Phase digram of the Bose-Hubbard model with pairing term. In (a) we fixed $\eta/U=0.025$. The phase diagram is presented for $\mu/U\in[1.5,3]$ and $J/U\in[0,0.0667]$. There are in total 4 types of phases. I for Mott insulator, II for non-chiral superfluid B, III for chiral ordered superfluid B and IV for chiral ordered superfluid A. In (b), $\eta/U=0.06$, there is one phase more, V stands for non-chiral superfluid A. }
\label{PhaseDia}
\end{figure}

One can find that there is a symmetry for three sites permutation in the unit cell. Therefore $\hat{\varphi}_+$ mode should be equivalent to $(-1,2,-1)/\sqrt{6}$ and $(-1,-1,2)/\sqrt{6}$. For simplicity, let us first neglect this $Z_3$ permutation symmetry and we will come back to this point later.  In the mean-field theory, we find the superfluid solution can be either $\varphi^R\neq 0$, $\varphi^I=0$ (superfluid A) or $\varphi^I\neq 0$, $\varphi^R=0$ (superfluid B). In Fig. \ref{Order}(a), we show the order parameters of $\varphi^R$ in superfluid A phase. In this state, $\Delta_+^R$ and $\Delta_0^R$ become nonzero first. When $\Delta_-^R=0$, time-reversal symmetry is not broken. While $\Delta_-^R$ becomes nonzero, $\varphi^R$ is complex, then chiral symmetry is broken. In Fig.\ref{Order}(b) we show the order parameters of $\varphi^I$ in the superfluid B phase. In this state, $\Delta_-^I$ emerges first while $\Delta_+^I$ and $\Delta_0^I$ follow. Finally, in Fig.\ref{Order}(c), we show the energy difference between two superfluid states. The lower energy state of these two superfluids is the ground state of the system. Therefore the system goes through a second-order transition when the order parameters become nonzero within a single state. Later we will find the level crossing transition between two superfluids is special and an enlarged symmetry emerges at the critical point.

To be more specific, we explain how these phase transitions happen following Fig. \ref{Order}, which is along the arrow line in the phase diagram given in Fig.\ref{PhaseDia} (b). 1) Mott insulator phase with all superfluid order parameters ${\boldsymbol\varphi}=0$ (phase I). The density is homogeneous in this phase.  2) As $J/U$ increases, the system goes through a superfluid Mott insulator transition to $\hat{\varphi}_-$ mode with only $\Delta_-^I\neq 0$. In this phase, the system is in a superfluid phase with a density order in $\hat{\varphi}_-$ mode (phase II). As mode $(1,-1,0)/\sqrt{2}$ and $(1,0,-1)/\sqrt{2}$ are equivalent to $\hat{\varphi}_-$ mode and the system has a parity symmetry in $\Delta_-^I$. Together the symmetry is $Z_3\times Z_2=Z_6$. The case we show here is one possibility. 3) As $J/U$ further increases, a level crossing happens, the condensation rotates from $\hat{\varphi}_-$ mode to a combination of $\hat{\varphi}_0$ and $\hat{\varphi}_+$ mode (phase V). This is another density-ordered superfluid phase. 4) An even larger $J/U$ induces a chiral symmetry broken, all three order parameters of ${\boldsymbol \varphi}^R$ are nonzero (phase IV). This phase is a chiral ordered superfluid.

Following the calculations for the order parameters, we can then draw a phase diagram for fixed $\eta/U$, varying $\mu/U$ and $J/U$. The total phase diagram is shown in Fig. \ref{PhaseDia}(a) and (b). Compared with the phase diagram at $\eta/U=0$, there are inhomogeneous density ordered superfluid  phases with, or without chiral order. It shows that a density-ordered superfluid is possible under the competition between frustration and the local interactions.

{\color{blue}\emph{Mechanism Study}}
Now we try to understand the phase transitions and figure out the mechanism. Here we treat $\eta$ term as a small perturbation. Based on Eq.~\ref{Eq:Decomp} , we find up to the second order of ${\boldsymbol\varphi}$, the ground state energy $E_g$ can be expressed as
\ba
E_g^{(2)}\!\!&=&\!\!\!\sum_{\sigma=\pm}\chi_\sigma\left(1-6J\chi_\sigma\right)\left((\Delta_\sigma^{R})^2+(\Delta_{-\sigma}^{I})^2\right)+\nonumber\\
&&\!\!\!\chi_+\left(1+6J\chi_+\right)(\Delta_0^{R})^2+\chi_-\left(1+6J\chi_-\right)(\Delta_0^{I})^2\label{Eq:MFT}
\ea
where $\chi_\pm=\chi\pm 2\eta\chi'$, and
\ba
\chi&=&\frac{\ell+1}{\epsilon_{\ell+1}-\epsilon_{\ell}}+\frac{\ell}{\epsilon_{\ell-1}-\epsilon_{\ell}},\\
\chi'&=&\frac{(\ell+1)(\ell+2)}{(\epsilon_\ell-\epsilon_{\ell+1})(\epsilon_\ell-\epsilon_{\ell+2})}+\frac{\ell(\ell+1)}{(\epsilon_\ell-\epsilon_{\ell+1})(\epsilon_\ell-\epsilon_{\ell-1})}\nonumber\\
&&+\frac{\ell(\ell-1)}{(\epsilon_\ell-\epsilon_{\ell-1})(\epsilon_\ell-\epsilon_{\ell-2})}.
\ea
Here $\ell=\lfloor \mu/U\rfloor$ is the integer part of $\mu/U$. $\epsilon_\ell=-\mu \ell+U\ell(\ell-1)/2$. We find in general $\chi_-<\chi_+$ as long as $\eta>0$. Therefore there are three typical regions for hopping strength $J$, that is, region I ($\chi_-<\chi_+<1/6J$), region II ($\chi_-<1/6J<\chi_+$), and region III ($1/6J<\chi_-<\chi_+$). It is then clear that in region I, all order parameters are zero, which corresponds to the MI phase. In region II, only $\Delta_+^R$ or $\Delta_-^I$ can be nonzero, where a density order related to $\hat{\varphi}_+$ and $\hat{\varphi}_-$ mode is spontaneously generated. Meanwhile, a superfluid order is developed with nonzero ${\boldsymbol\varphi}$.  Finally, in region III, the order parameter becomes complex such that a chiral order is developed. 

Further, we will only focus on the MI-DWSF transition and the DWSF transition within region II. As we are focusing on region I and region II, only order parameters $\Delta_-^I$, $\Delta_+^R$ and $\Delta_0^R$ are important. To better describe our problem, we introduce real order parameter $\varphi\in {\mathbb R}$, $\theta\in S^1$, and $\phi\in{\mathbb R}$, where $\Delta_-^I=\varphi\cos\theta$, $\Delta_+^R=\varphi\sin\theta$ and $\Delta_0^R=\phi$. Then we find around the MI-DWSF transition point, 
\ba
E_g\approx u_2\varphi^2+u_4\varphi^4+u_6\varphi^6-u_6^\prime \varphi^6 \cos 6\theta
\ea
where $u_2$, $u_4$, $u_6$, $u_6^\prime$ are phenomenological parameters \cite{supp}.
At zero temperature, the MI-DWSF transition is characterized by sign change in $u_2$ with $u_4>0$, $u_6>0$ and $u_6^\prime>0$. Here as $u_6^\prime >0$, therefore $\Delta_-^I\neq 0$, $\Delta_+^R=0$ is a solution. 

\begin{figure}[t]
\includegraphics[width=8.8cm]{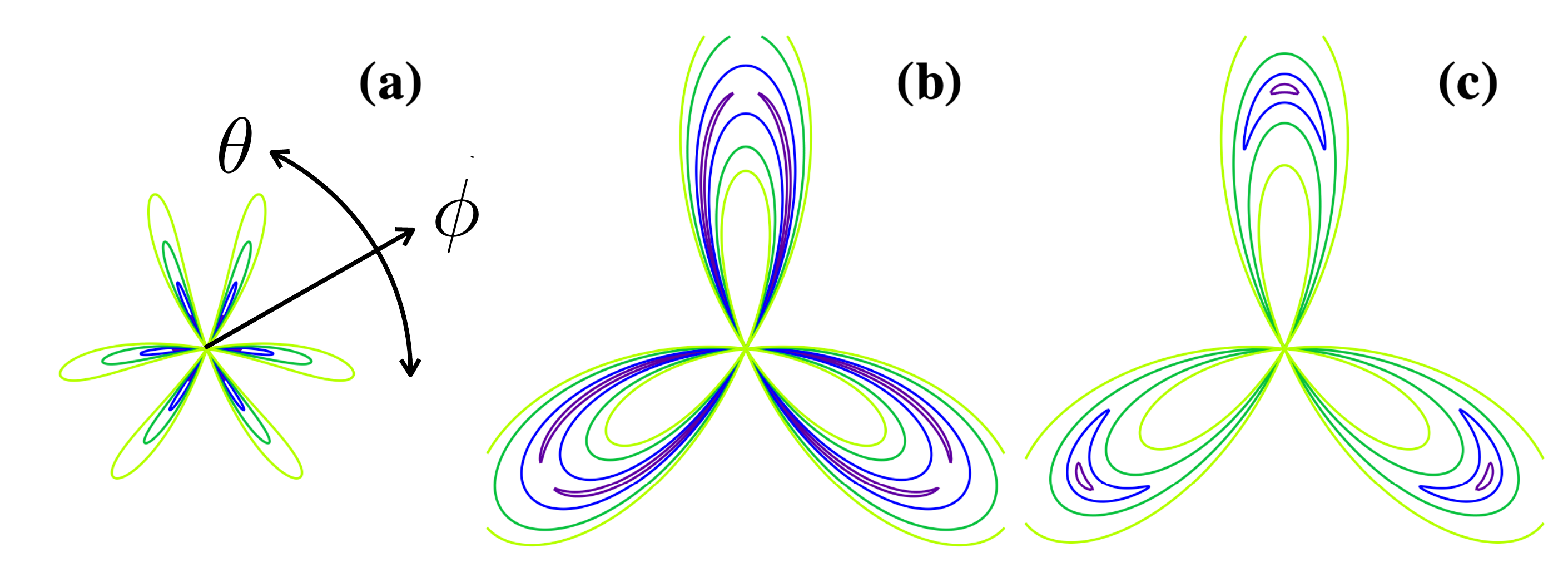}
\caption{Cross sections of the energy curve for Eq. \ref{SixClock}. For small $\lambda$, the typical energy cross section is presented in (a). The minimum is surrounded by the purple curve. We can see the minima are located at $\phi=0$, $\theta=n\pi/3$, $n\in{\mathbb Z}$. For large $\lambda$, the superfluid phase change in $\Delta_+^R\neq 0$, $\Delta_0^R\neq 0$ case. The typical energy cross section is shown in (c). Here $\phi>0$, $\phi\neq 0$ and $\theta=3\pi/6$, $7\pi/6$ and $11\pi/6$. In (b), we have shown the energy cross section of a $\lambda$ close to critical point. We can see there is a purple curve shaped like a shamrock, where the ground state energy are almost degenerate along the purple line. This is the case for emergent U(1) symmetry. }\label{ClockTransition}
\end{figure}

In region II, let us assume that the superfluid order is large enough that we can neglect the fluctuations of the magnitude mode of $\varphi$, then we find the effective ground state energy can be written as\cite{supp} 
\ba
E_g=-u\cos6\theta+\lambda\phi\sin3\theta+m^2\phi^2\label{SixClock}
\ea
When $u>0$, $\theta=2n\pi/6$, $n=0,1,2,3,4,5$ are ground state manifold solutions. This solution is just $\Delta_-^I\neq 0$ DWSF, a condensation in $\hat{\varphi}_-$ mode. On the other hand, when $u<\lambda^2/8m^2$, $\phi=-\lambda\sin3\theta/2m^2$, we should take $\cos 6\theta=-1$, where $\theta=(2n+1)\pi/6$, such that $\sin3\theta=\pm1$. This is the other metastable solution in our calculation as $\Delta_+^R\neq 0$, $\Delta_0^R\neq 0$. The picture for this phase transition is shown in Fig.\ref{ClockTransition} (a)-(c), where $\phi$ is put in polar direction and $\theta$ is put in angle direction. One can see in Fig.\ref{ClockTransition} (a), the ground state takes $\phi=0$ and $\theta= n\pi/3$.  As $\lambda$ increases, in Fig.\ref{ClockTransition}, the vacuum becomes $\phi>0$, and $\theta=\frac{\pi}{2},\frac{7\pi}{6}, \frac{11\pi}{6}$ or $\phi<0$, $\theta=\frac{\pi}{6},\frac{5\pi}{6}, \frac{3\pi}{2}$. One can observe from Fig.\ref{ClockTransition}(b) or from Eq.\ref{SixClock} that at the transition point, the ground state energy does not have $\theta$ dependence, therefore a U(1) symmetry is emergent at the transition point. As we see the energy minima lies in a shamrock-like curve. Therefore we call the transition ``shamrock transition". Although the order parameter jumps at the transition point, the shamrock transition is not a first-order transition. The transition is between two symmetry broken states, no symmetry is broken across the transition. Therefore this is neither a spontaneously symmetry broken transition. For this reason, the shamrock transition is beyond the Landau-Ginzberg paradigm. 

The effective theory given in Eq.~\ref{SixClock} is in the same universality class of the six-state clock model \cite{clock6_77,clock6_86,clockmodel,Sandvik2021,MengQi2021}, and it enables us to have some guesses for finite temperature phases. For $d=2$ ($d$ is the spatial dimension) case, if the system doesn't have particle-hole symmetry, then the total dimension $D$ is still 2 ($D$ is the quantum dimension). Then when the temperature rises, the system first goes through a first-order transition from DWSF to a KT phase where U(1) symmetry is recovered. As the temperature continues to increase, the KT superfluid phase becomes a disordered phase. Although it may not be so surprising from an effective field theory view, the U(1) symmetry here is quite different from the original U(1) symmetry. Under this U(1) symmetry the density pattern can rotate from (0,1,-1) mode to (2,-1,-1) mode, which seems to be a very exotic non-homogeneous superfluid.  While if there is particle-hole symmetry, the $(\partial_\tau\theta)^2$ term becomes leading order in the effective action, therefore the total dimension becomes 3. In 3 dimensional space, there is only one transition from ordered phase to disordered phase. Therefore we expect a critical line starts from the quantum critical point.

{\color{blue}\emph{Conclusion and Outlook}}  To summarize, we have proved that a density wave ordered superfluid is possible due to the competition between frustration and local interactions. In the modified Bose-Hubbard model on a triangle lattice, a Mott insulator phase and four density wave ordered superfluid phases are discovered by the mean-field theory. An interesting transition (the shamrock transition) between two non-chiral DWSFs is discovered where an emergent U(1) symmetry is presented at the critical point while both phases on each side have a $Z_6$ symmetry. This model can be realized in Kerr cavities or exciton-polariton systems. It is a good platform for density-ordered superfluid studies and KT transition studies. Different from frustrated spin systems, the Higgs modes are unlocked in the present system. It is also a good platform to study Goldstones with light masses. Many new physics are generated due to the unlocked amplitude fluctuations.

{\color{blue}\emph{Acknowledgements}}  We would like to thank Shaokai Jian and Shuai Yin for valuable discussions. Y. C is supported by Beijing Natural Science Foundation (Z180013), and NSFC under Grant No. 11734010. C. W is supported by the Beijing Outstanding Young Scientist Program, NSFC Grant No. 11734010, MOST Grant No. 2016YFA0301600.

\appendix

\setcounter{equation}{0}
\renewcommand{\theequation}{S\arabic{equation}}

\begin{widetext}
\newpage
\centerline{\bf Supplementary Material}

\section{Effective Theory at $\varphi$'s second order and $\eta$'s first order}
The mean field hamiltonian is 
\ba
\hat{H}_{\rm MF}&=&\sum_{j=1,2,3}\hat{H}_{0,j}+\hat{V}_j\\
\hat{H}_{0,j}&=&-\mu\hat{n}_j+\frac{U}{2}\hat{n}_j(\hat{n}_j-1)\\
\hat{V}_j&=&\eta(\hat{a}^{}_j\hat{a}^{}_j+\hat{a}^\dag_j\hat{a}^\dag_j)+(\varphi_{j-1}+\varphi_{j+1})\hat{a}^\dag_j+(\varphi^*_{j-1}+\varphi^*_{j+1})\hat{a}_j^{}
\ea
Upto $\varphi$'s second order and $\eta$'s first order, the ground state wave function is

\ba
|\Psi\rangle_j=\frac{1}{\sqrt{N_\psi}}\left(|\ell\rangle+\sum_{\ell_1}\frac{|\ell_1\rangle\langle\ell_1|\hat{V}_j|\ell\rangle}{\epsilon_\ell-\epsilon_{\ell_1}}+\sum_{\ell_1,\ell_2}\frac{|\ell_2\rangle\langle\ell_2|\hat{V}_j|\ell_1\rangle\langle\ell_1|\hat{V}_j|\ell\rangle}{(\epsilon_{\ell}-\epsilon_{\ell_1})(\epsilon_{\ell}-\epsilon_{\ell_2})}\right).
\ea
More explicitly, $|\Psi\rangle=\prod_{j=1,2,3}\otimes|\Psi\rangle_j$ is a product state of each site's wave function, where 
\ba
|\Psi\rangle_j&=&|\ell\rangle+\frac{\sqrt{\ell+1}\varphi}{\epsilon_\ell-\epsilon_{\ell+1}}|\ell+1\rangle+\frac{\sqrt{\ell}\varphi^*}{\epsilon_\ell-\epsilon_{\ell-1}}|\ell-1\rangle+\frac{\eta\sqrt{(\ell+2)(\ell+1)}}{\epsilon_\ell-\epsilon_{\ell+2}}|\ell+2\rangle+\frac{\eta\sqrt{\ell(\ell-1)}}{\epsilon_\ell-\epsilon_{\ell-2}}|\ell-2\rangle+\nonumber\\
&&\frac{\sqrt{(\ell+1)(\ell+2)}\varphi^2}{(\epsilon_\ell-\epsilon_{\ell+1})(\epsilon_\ell-\epsilon_{\ell+2})}|\ell+2\rangle+\frac{\sqrt{\ell(\ell-1)}\varphi^{*2}}{(\epsilon_\ell-\epsilon_{\ell-1})(\epsilon_\ell-\epsilon_{\ell-2})}|\ell-2\rangle+\frac{\eta\varphi(\ell+1)\sqrt{\ell}}{(\epsilon_\ell-\epsilon_{\ell+1})(\epsilon_\ell-\epsilon_{\ell-1})}|\ell-1\rangle+\nonumber\\
&&\frac{\eta\varphi^*\ell\sqrt{\ell+1}}{(\epsilon_\ell-\epsilon_{\ell+1})(\epsilon_\ell-\epsilon_{\ell-1})}|\ell+1\rangle+\frac{\eta\sqrt{\ell}(\ell-1)\varphi}{(\epsilon_\ell-\epsilon_{\ell-1})(\epsilon_{\ell}-\epsilon_{\ell-2})}|\ell-1\rangle+\frac{\eta\varphi^*(\ell+2)\sqrt{\ell+1}}{(\epsilon_\ell-\epsilon_{\ell+1})(\epsilon_{\ell}-\epsilon_{\ell+2})}|\ell+1\rangle.\ea
Here $\varphi$ is short for $\varphi_{j-1}+\varphi_{j+1}$.
With the help of the perturbative ground state wave function, the ground state energy can be expressed as
\ba
E_g&=&\langle\Psi|\hat{H}|\Psi\rangle=\langle\Psi|\hat{H}_{\rm MF}-\sum_j\left((\varphi_{j-1}+\varphi_{j+1})\hat{a}_j^\dag+h.c.\right)+3J\sum_j(\hat{a}_{j+1}^\dag\hat{a}^{}_j+h.c.)|\Psi\rangle\nonumber\\
&=&E_g^{\rm MF}+\sum_j(3J\langle\hat{a}_{j-1}\rangle+3J\langle\hat{a}_{j+1}\rangle-\varphi_{j-1}-\varphi_{j+1})\langle\hat{a}_j^\dag\rangle+h.c.),
\ea
where $E_g^{\rm MF}$ is the mean field ground state energy $\langle \Psi|\hat{H}_{\rm MF}|\Psi\rangle$, and $\langle\hat{a}_j\rangle\equiv\langle\Psi|\hat{a}_j|\Psi\rangle$. Again the summation over $j$ is cyclic (the summation and subtraction is module 3 in a sense 3+1=1). More explicitly, we have
\ba
E_g^{\rm MF}&=&-\chi\sum_j|\varphi_{j-1}+\varphi_{j+1}|^2-\eta\chi'\sum_j\left((\varphi_{j-1}+\varphi_{j+1})^2+c.c.\right),
\ea
and
\ba
\langle\hat{a}_j\rangle=-\chi(\varphi_{j-1}+\varphi_{j+1})-2\eta\chi'(\varphi_{j-1}^*+\varphi_{j+1}^*)+{\cal O}(\varphi^3),
\ea
where
\ba
\chi&=&-\frac{\ell+1}{\epsilon_\ell-\epsilon_{\ell+1}}-\frac{\ell}{\epsilon_\ell-\epsilon_{\ell-1}}\\
\chi'&=&-\frac{\ell+1}{(\epsilon_\ell-\epsilon_{\ell+1})}\frac{\ell+2}{\epsilon_\ell-\epsilon_{\ell+2}}-\frac{\ell+1}{(\epsilon_\ell-\epsilon_{\ell+1})}\frac{\ell}{\epsilon_\ell-\epsilon_{\ell-1}}-\frac{\ell}{(\epsilon_\ell-\epsilon_{\ell-1})}\frac{\ell-1}{\epsilon_\ell-\epsilon_{\ell-2}}
\ea
By introducing $\varphi_1=\frac{1}{\sqrt{3}}(\Delta_0^R+i\Delta_0^I)+\frac{2}{\sqrt{6}}(\Delta_+^R+i\Delta_+^I)$, $\varphi_2=\frac{1}{\sqrt{3}}(\Delta_0^R+i\Delta_0^I)-\frac{1}{\sqrt{6}}(\Delta_+^R+i\Delta_+^I)+\frac{1}{\sqrt{2}}(\Delta_-^R+i\Delta_-^I)$, and $\varphi_3=\frac{1}{\sqrt{3}}(\Delta_0^R+i\Delta_0^I)-\frac{1}{\sqrt{6}}(\Delta_+^R+i\Delta_+^I)-\frac{1}{\sqrt{2}}(\Delta_-^R+i\Delta_-^I)$, we have
\ba
E_g=\sum_{\sigma=\pm}\chi_\sigma\left(1-6J\chi_\sigma\right)\left((\Delta_\sigma^{R})^2+(\Delta_{-\sigma}^{I})^2\right)+\chi_+\left(1+6J\chi_+\right)(\Delta_0^{R})^2+\chi_-\left(1+6J\chi_-\right)(\Delta_0^{I})^2,\ea
where $\chi_\pm=\chi\pm 2\eta\chi'$. This is the Equation 8 in the main text.

\section{The Effective Theory With Real Order Parameters Only}

In some parameter region, the order parameters are not complex. The presence of complex order means chiral symmetry broken. Therefore when we restricted ourselves in non-chiral superfluid, we can simplify the effective field theory. 

In mean field theory, we assume the on-site hamiltonian as 
\ba
\hat{H}_j=-\mu\hat{n}_j+\frac{U}{2}\hat{n}_j(\hat{n}_j-1)-\eta(\hat{a}^{}\hat{a}^{}+\hat{a}^\dag\hat{a}^\dag)+\hat{a}(\varphi_{j-1}+\varphi_{j+1})^*+\hat{a}^\dag(\varphi_{j-1}+\varphi_{j+1})
\ea
Here we assume that the parameters are restricted to a phase where $\varphi_{j=1,2,3}$ are all real.

First we are going to prove that if the mea field on-site energy $E_{g, {\rm on-site}}^{\rm MF}\equiv \sum_j\langle \Psi_g|\hat{H}_j-\hat{a}_j^\dag(\varphi_{j-1}+\varphi_{j+1})-\hat{a}_j(\varphi_{j-1}+\varphi_{j+1})^*|\Psi_g\rangle$ can be written in the following form,
\ba
E_{g, {\rm on-site}}^{\rm MF}=-\sum_{n=1}^3\sum_{j} r_n(\varphi_{j-1}+\varphi_{j+1})^{2n}+{\cal O}(\varphi^8),\label{EqOnsiteE}
\ea
Then the average of $\hat{a}_j$ must be
\ba
\langle \hat{a}_j\rangle=\sum_{n=1}^{3} (n r_n)(\varphi_{j-1}+\varphi_{j+1})^{2n-1}+{\cal O}(\varphi^7)\label{EqAlpha}
\ea
This is the result of $\partial(\sum_j \langle\hat{H}_j\rangle)/\partial(\varphi_{j-1}+\varphi_{j+1})=0$.  Further, we find the true ground state energy $E_g=\langle \hat{H}\rangle$. In mean field approximation, it is
\ba
E_g&=&E_{g, {\rm on-site}}^{\rm MF}+E_{g,{\rm kin}},\\
E_{g,{\rm kin}}&\equiv&\langle\Psi_g| J\sum_{\langle i,j\rangle}(\hat{a}_i^\dag\hat{a}_j^{}+\hat{a}_j^\dag\hat{a}_i^{})|\Psi_g\rangle\approx J\sum_{\langle i,j\rangle } (\langle \hat{a}_i^\dag\rangle\langle \hat{a}_j\rangle+c.c.)\label{Energy}
\ea
By inserting Eq.~\ref{EqOnsiteE} and Eq.~\ref{EqAlpha} into Eq.~\ref{Energy}, and we introduce
\ba
{\boldsymbol \varphi}=(\varphi_1,\varphi_2,\varphi_3)\equiv\varphi \sin\theta\hat{\varphi}_++\varphi\cos\theta\hat{\varphi}_-+\phi\hat{\varphi}_0,
\ea
where $\hat{\varphi}_0=(1,1,1)/\sqrt{3}$, $\hat{\varphi}_+=(2,-1,-1)/\sqrt{6}$, and $\hat{\varphi}_-=(0,1,-1)/\sqrt{2}$, we have
\ba
E_g&=&(r_1-\tilde{J}r_1^2)\varphi^2+(2\tilde{J} r_1 r_2-\frac{1}{2}r_2)\varphi^4+\frac{1}{36}(-28\tilde{J}r_2^2+10r_3-60\tilde{J}r_1 r_3)\varphi^6+\frac{1}{36}(-8\tilde{J}r_2^2-r_3+6\tilde{J}r_1r_3)\varphi^6 \cos(6\theta)+\nonumber\\
&&\frac{4\sqrt{2}}{3}(-r_2+\tilde{J}r_1r_2)\sin(3\theta)\varphi^3 \phi+\sqrt{2}\left(\frac{4\tilde{J}}{3}r_2^2+\frac{5}{3}r_3-5\tilde{J}r_1r_3\right)\sin(3\theta)\varphi^5\phi+\nonumber\\
&&(4r_1+8\tilde{J}r_1^2)\phi^2+(-8r_2-16\tilde{J}r_1r_2)\phi^2\varphi^2+(8\tilde{J}r_2^2+10r_3)\varphi^4\phi^2
\ea
where $\tilde{J}=6J$. Here to simplify the effective theory, we find the second order coefficient of $\phi$ field is always positive. Initially, when we consider the superfluid transition, we fix $\phi=0$, then the effective theory is
\ba
E_g= u_2\varphi^2+u_4\varphi^4+u_6\varphi^6-u_6^\prime \varphi^6\cos(6\theta),
\ea
where $u_2=r_1-\tilde{J}r_1^2$, $u_4=2\tilde{J} r_1 r_2-\frac{1}{2}r_2$, $u_6=(-28\tilde{J}r_2^2+10r_3-60\tilde{J}r_1 r_3)/36$, and $u'_6=(-8\tilde{J}r_2^2-r_3+6\tilde{J}r_1r_3)/36$. Within the superfluid phase, if we assume $\varphi\neq 0$ is fixed, then
\ba
E_g=-u\cos(6\theta)+\lambda\phi\sin3\theta+m^2\phi^2,
\ea
where $u$, $m^2$, $\lambda$ are all $\varphi$ dependent parameters. These parameters can be obtained when $r_{n=1,2,3}$ and $\varphi$ are given. 

\end{widetext}

\end{document}